\documentclass[%
 aip,
 cha,
 sd,%
 amsmath,amssymb,
 reprint,%
]{revtex4-1}

\usepackage{amsmath}	% required for `\align' (yatex added)

\newcommand{\av}[1]{\langle{#1}\rangle}
\usepackage[normalem]{ulem}

\usepackage{here}
\usepackage[dvipdfmx]{graphicx}% Include figure files
\usepackage{dcolumn}% Align table columns on decimal point
\usepackage{bm}% bold math

\begin{document}

\preprint{AIP/123-QED}

\title[Synchronization failure caused by interplay between noise and network
heterogeneity]{Synchronization failure caused by interplay between noise and network heterogeneity}% Force line breaks with \\

\author{Y. Kobayashi}
 \email{yasuaki.kobayashi@es.hokudai.ac.jp}
 \affiliation{
Research Institute for Electronic Science, Hokkaido University, Sapporo, 060-0811, Japan
 %\\This line break forced with \textbackslash\textbackslash
}%
% \altaffiliation[Also at ]{Physics Department, XYZ University.}%Lines break automatically or can be
  % forced with \\
  
  \author{H. Kori}%
   \email{kori.hiroshi@ocha.ac.jp}
\affiliation{%
Department of Information Sciences, Ochanomizu University, Tokyo, 112-8610, Japan%\\This line break forced% with \\
}%

\date{\today}% It is always \today, today,
	     
\begin{abstract}
We investigate synchronization in complex networks of noisy phase oscillators. We find that, while
 too weak a coupling is not sufficient for the whole system to synchronize, too strong a coupling
 induces a nontrivial type of phase slip among oscillators, resulting in synchronization
 failure. Thus, an intermediate coupling range for synchronization exists, which becomes narrower when the network is more heterogeneous. Analyses of two noisy oscillators reveal that nontrivial phase slip
 is a generic phenomenon when noise is present and coupling is strong. Therefore, the low
 synchronizability of heterogeneous networks can be understood as a result of the difference in effective coupling strength among oscillators with different degrees; oscillators with high degrees tend to
 undergo phase slip while those with low degrees have weak coupling strengths that are insufficient for synchronization.
 \end{abstract}

\pacs{}% PACS, the Physics and Astronomy
\keywords{synchronization, heterogeneous network, noise}%Use showkeys class option if keyword
\maketitle

\begin{quotation}
  Synchronization phenomena are found in various systems, where the maintenance of synchronization is quite often crucial for proper functioning. Such systems are represented by mutually
 interacting oscillators, and their collective dynamics depend both on the interaction function
 between a pair of connected oscillators and on the network structure among the oscillators. Here, we
 show that when such coupled oscillators are under the influence of noise, too strong a coupling induces nontrivial phase slip. Moreover, we show that heterogeneous networks, which have a
 wide dispersion of network connections among individual oscillators, are more strongly affected by
 the nontrivial phase slip.  While synchronization failure is known to occur in heterogeneous networks of a particular
 class of chaotic oscillators, our study demonstrates the difficulty of synchronization in
 heterogeneous networks of periodic oscillators under the influence of noise. 
\end{quotation}

\section{Introduction}
Synchronized oscillation of active elements can be observed in various fields, including biology,
engineering, ecosystem, and chemical systems \cite{winfree01, kuramoto84, pikovsky01, glass01, kiss07}.
In many cases, synchronization of the entire system is required for
proper functioning under various types of noise and heterogeneity.
Important examples include the heart (a population of cardiac cells) \cite{tweel,glass01}, the circadian
pacemaker (a population of clock cells) \cite{reppert02,yamaguchi03}, and power grids \cite{motter2013,doerfler,rohden}. 

Oscillators are often connected through a complex network, where heterogeneity
in network connectivity may critically hamper synchronization.
Such a case is actually observed in a special type of
chaotic oscillators. For the synchronization of these chaotic oscillators \cite{fujisaka}, there
exist both lower and upper thresholds of the coupling strength \cite{heagy},
and thus, global synchronization tends to fail for heterogeneous networks \cite{nishikawa03}.
However, only a few examples of such chaotic oscillators with
a similar property are known. Moreover, for periodic oscillators,
little has been reported about the potential negative effect of network heterogeneity
on synchronization. 

We have recently investigated a phase oscillator that is unilaterally influenced by a pacemaker
and is also under noise \cite{koba_kori2015} and found nontrivial phase slip in the strong coupling
regime; this implies that synchronization is possible only for intermediate coupling strengths. It
has been shown that this reentrant transition is observed for general interaction
functions. Although it is not at all obvious whether dynamical behavior obtained in the phase
model with strong coupling is also reproduced in limit-cycle oscillators, it has been demonstrated
that the Brusselator model, a typical system of limit cycle oscillators, does show the reentrant transition. Such nontrivial phase slip can be another source of instability for a
population of network-coupled oscillators, and thus it is important to understand how different
networks respond to this new type of instability. 

In this study, we show that heterogeneous networks are more prone to phase slip. The essence of this behavior can be understood from the phase diagram of two mutually coupled phase oscillators.
Mutually coupled oscillators fail to synchronize for too weak and too strong coupling strengths
when they are subjected to noise. Because of this property, synchronization is easily violated in
heterogeneous networks.

This paper is organized as follows: in Sec.~\ref{sec:mutual}, we investigate a system of two mutually
interacting oscillators under noise and compare it with two unilaterally connected oscillators, which have been studied previously; two main
causes of frequency drop, nontrivial phase slip and oscillation death, are discussed
in Sec.~\ref{sec:slip} and Sec.~\ref{sec:death}, respectively. Then, in
Sec.~\ref{sec:network} we present the results for network-connected
oscillators. 

\section{Two mutually interacting oscillators}\label{sec:mutual}
Let us first consider two mutually interacting identical phase oscillators subjected to noise. Their
dynamics are governed by the following phase equation:
\begin{equation}
 \dot{\phi}_i = \omega + K Z(\phi_i)\left\{h(\phi_j)-h(\phi_i)\right\}+\xi_i,\label{eq:model}
\end{equation}
where $(i,j)=(1,2)$ or $(2,1)$, $\phi_i(t)$ and $\omega$ are the phase
and the natural frequency of the $i$-th
oscillator, respectively, $K>0$ is the coupling strength, and $\xi_{1,2}(t)$ is Gaussian white noise with
strength $D$, \textit{i.e.}, $\langle\xi_i(t)\xi_j(t') \rangle=D\delta_{ij}\delta(t-t')$.
We set $\omega=1$ without loss of generality. 
This model is symmetrical
under oscillator exchange, and interaction vanishes when $\phi_1=\phi_2$. The precise form
of the interaction is determined by the phase sensitivity function $Z(\phi)$ and the stimulus function
$h(\phi)$ \cite{winfree67, winfree01}, which are both $2\pi$-periodic functions. We specifically choose
\begin{equation}
 Z(\phi)=\sin(\phi-\alpha), \quad h(\phi)=-\cos\phi, \label{eq:functions}
\end{equation}
where $\alpha$ is a parameter. Throughout this work we assume $|\alpha|<\frac{\pi}{2}$, which
assures that the synchronous state is linearly stable in the absence of noise. 

When $K$ is small compared to $\omega$, the averaging approximation is
applicable to Eq.~\eqref{eq:model} \cite{kuramoto84, hoppensteadt97}, resulting in
\begin{equation}
 Z(\phi_i)\left\{h(\phi_j)-h(\phi_i)\right\} \approx
  \frac{1}{2}\{\sin(\phi_j - \phi_i + \alpha) - \sin \alpha\},\label{eq:averaged}
\end{equation}
which is the
Sakaguchi-Kuramoto coupling function \cite{sakaguchi_kuramoto, koba_kori2015}. We refer to
Eq.~\eqref{eq:model} as the non-averaged phase model and to the same one with the approximated interaction given by Eq.~\eqref{eq:averaged} as the averaged phase model. Note that the averaged phase model is valid
as a model of coupled limit-cycle oscillators only for $K\ll \omega$. Since we are
concerned with both weak and strong coupling strengths we employ the non-averaged
phase model in this work. Although considering such non-averaged cases would normally require us to treat multiplicative
noise proportional to $Z(\phi)$, here we consider additive noise for simplicity. 

In a previous study \cite{koba_kori2015} we have investigated a phase oscillator with phase $\phi$
under noise that is unilaterally coupled to a noise-free pacemaker with the same oscillation frequency $\omega$:
\begin{equation}
 \dot{\phi} = \omega + K Z(\phi)\left\{h(\omega t)-h(\phi)\right\}+\xi, \label{eq:previous}
\end{equation}
where the functions $Z$ and $h$ are the same as in Eq.~\eqref{eq:functions}, and equally $\xi$ is
the same Gaussian white noise. In this unilateral model, we found the following reentrant
transition: as $K$ increases from zero with a fixed value of $D$, the oscillator undergoes the first
transition from a noise-driven to a pacemaker-driven synchronous state; and then, as $K$ increases
further the oscillator undergoes a second transition, after which phase slip occurs more frequently
with increasing $K$. Before the first transition, diffusion causes phase slip, where the effect of noise
is stronger than the effect of coupling; this type of phase slip is trivial and also occurs in the
averaged phase model, such as for Kuramoto or Sakaguchi-Kuramoto oscillators. In contrast, phase
slip after the second transition is counter-intuitive in the sense that stronger coupling yields
more frequent synchronization failure and that at each slip event the oscillator lags behind the pacemaker
(i.e., the phase difference $\phi-\omega t$ decreases by $2\pi$). It has been shown that this
nontrivial phase slip is caused by an interplay between noise and nonlinearity and that this occurs only in non-averaged phase models with noise \cite{koba_kori2015}.

\begin{figure*}[htb]
\includegraphics[width=\linewidth]{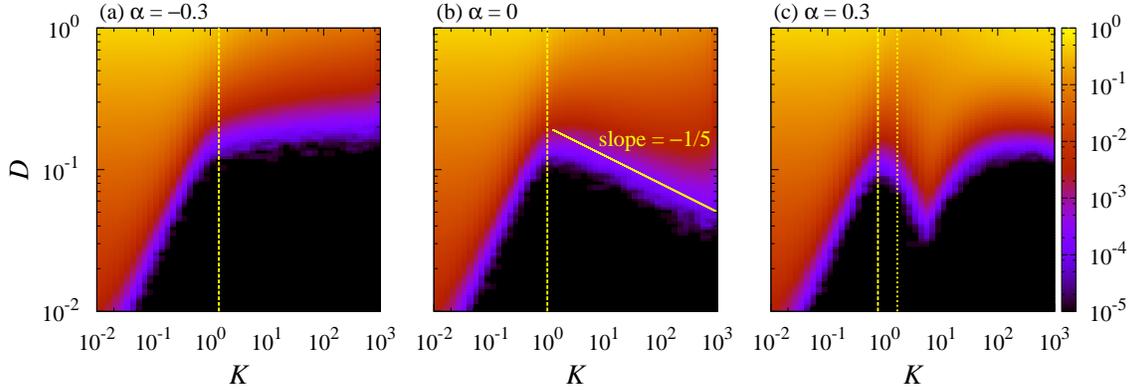}
 \caption{Phase slip rate for Eq.~\eqref{eq:model} as a function of the coupling strength $K$ and the
 noise strength $D$ for three different values of model parameter $\alpha$: (a) $\alpha=-0.3$,
 (b) $\alpha=0$, and (c) $\alpha=0.3$. In (b), the slope of the right boundary between the synchronous state
 and the phase slip state is theoretically given by Eq.~\eqref{eq:scaling}. Vertical dashed lines in
 (a) and (b) represent $K=K_{c1}$, where $K_{c1}$ is defined in Eq.~\eqref{eq:kc1}, and two vertical lines in (c) represent $K=K_{c1}$ and $K=K_{c2} > K_{c1}$, respectively, where
 $K_{c2}$ is defined in Eq.~\eqref{eq:kc2}. }\label{fig:phasediag_mutual}
\end{figure*}
Phase slip is also observed in the present model \eqref{eq:model} where interaction is mutual. A
phase slip event is counted every time when the phase difference $\psi\equiv\phi_1-\phi_2$ increases or decreases by
$2\pi$, and the slip rate is defined as the total number of phase slips divided by the observation
time $t_{\mathrm{max}}=10^5$.

\begin{figure}[htb]
\includegraphics[width=\linewidth]{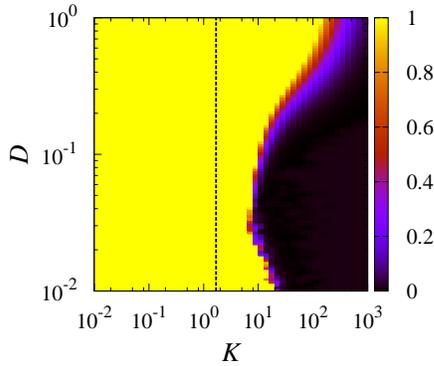}
 \caption{Mean oscillation frequency of the two oscillators in Eq.~\eqref{eq:model}, averaged
 over observation time, as a function of the coupling strength $K$ and the noise intensity $D$ for $\alpha=0.3$, corresponding to Fig.~\ref{fig:phasediag_mutual}(c). The dashed vertical line represents $K=K_{c2}$ [Eq.~\eqref{eq:kc2}]. }\label{fig:death}
\end{figure}

Figure \ref{fig:phasediag_mutual} shows the slip rate as a function of the coupling strength $K$ and
the noise intensity $D$ for three different $\alpha$ values. The phase slip predicted by the
averaged model\cite{pikovsky01,koba_kori2015} is observed for all cases where $D>K$. In addition to this trivial type
of phase slip,  reentrant transitions are also observed for a range of $D$ when $\alpha\ge 0$. In
particular, for $\alpha=0$, the second transition line follows a power law with an exponent close to $-0.2$. For $\alpha=0.3$, the reentrant transition line has a steeper
slope. Also, for large $K$, the phase slip region is invaded by the region of oscillation death. Figure
\ref{fig:death} plots the mean oscillation frequency of the two oscillators, averaged over
observation time $t_{\mathrm{max}}$, with the average frequency for each oscillator given by $\av{\omega_i}\equiv
\frac{1}{t_{\mathrm{max}}}\int_0^{t_{\mathrm{max}}}\dot{\phi}_idt$. The black region indicates that
oscillation completely stops. No such oscillation death is observed for $\alpha \le 0$. For
$\alpha=-0.3$, the reentrant region disappears, although nontrivial phase slip is observed for
$D<K$. We have numerically confirmed that there are no reentrant regions observed for at least $\alpha<-0.1$ and that the region of nontrivial phase slip diminishes as $\alpha$ decreases.

The overall tendency of phase slip for different $\alpha$ is similar to our previous result for
unilateral coupling [Eq.~\eqref{eq:previous}] in that the model shows reentrant transitions and the nontrivial phase slip region expands as $\alpha$ increases, although there are differences too. In the case of
unilateral coupling, no oscillation death is observed, which is obvious because of the existence of the
pacemaker. Further, the power law exponent, which also appears in the case of unilateral coupling when
$\alpha=0$, is theoretically estimated and numerically confirmed\cite{koba_kori2015} to be
$-\frac{1}{3}$, which is far from the value of $-0.2$ observed in Fig.~\ref{fig:phasediag_mutual}(b). The difference between the exponents indicates
that the mechanism of phase slip differs from the case of unilateral interaction, which is investigated below.

\subsection{Nontrivial phase slip}\label{sec:slip}
Let us investigate how phase slip occurs for strong coupling. To do this, we rewrite model \eqref{eq:model} in terms of mean phase $\Phi \equiv \frac{\phi_1+\phi_2}{2}$ and phase difference
$\psi \equiv \phi_1-\phi_2$:
\begin{align}
 \dot{\Phi}&=\omega - K\sin^2\frac{\psi}{2}\{\sin(2\Phi-\alpha)+\sin\alpha\} + \xi', \label{eq:phi_model}\\
 \dot{\psi}&=K\{\cos(2\Phi-\alpha)-\cos\alpha\}\sin\psi + \xi'', \label{eq:psi_model}
\end{align}
where $\xi'$ and $\xi''$ are Gaussian white noise with the noise intensity $D/2$ and $2D$,
respectively. Note that the system is $\pi$-periodic in $\Phi$ and $2\pi$-periodic in $\psi$.

In the noise-free case, there exists a synchronous solution $(\Phi, \psi) = (\omega t, 0)$.
The synchronous state $\psi=0$ is linearly stable for $K>0$, with Floquer multiplier
$e^{-\pi K\cos\alpha}$. 
When a sufficiently strong noise is introduced, this synchronous state may be destabilized
according to the following scenario: if $K$ is large, phase difference $\psi$ is strongly bound to zero for most of the time during
the evolution of $\Phi$, but there are instances in which the coupling term for $\psi$ vanishes, namely
when $\Phi$ is close to $0$ and $\alpha$, which means that short time intervals exist in which $\psi$ is
driven only by noise. If, during this noise-driven period, the deviation of $\psi$ from the synchronous
state $\psi=0$ becomes large, the second term of Eq.~\eqref{eq:phi_model} becomes comparable to
the first term, leading to $\dot{\Phi}\sim 0$. Then, it is likely that $\psi$ continues to be driven by noise
with the mean phase velocity kept close to zero, until $\psi$ makes a full revolution. Hence, a
necessary condition for nontrivial phase slip is specified from the requirement that $\dot{\Phi}=0$ has a solution
in the noise-free case: since $\dot{\Phi} > \omega-K(1+\sin\alpha)$, $\dot{\Phi}=0$ is possible only if
\begin{align}
 K > K_{c1} \equiv \frac{\omega}{1+\sin\alpha}. \label{eq:kc1}
\end{align}
As shown in Fig.~\ref{fig:phasediag_mutual}, each phase diagram in
Fig.~\ref{fig:phasediag_mutual} is separated by a line $K=K_{c1}$, and the region $K<K_{c1}$
shows the behavior expected from the averaged phase model. 

In the case of $\alpha=0$, the following scaling analysis validates the above scenario. Let us first
estimate the range of $\Phi$ in which the dynamics of $\psi$ are governed only by noise. Below, the phase difference is considered in the range $0\le\psi<\pi$. Suppose that $\Phi=0$ at $t=0$. In this moment the
coupling term for $\psi$ vanishes. When $t$ is small, dynamics at around $t=0$ are governed at the lowest order of $\Phi$ by
\begin{align}
 \dot{\Phi}&=\omega-2K \left(\sin^2\frac{\psi}{2}\right)\Phi+\xi',\\
 \dot{\psi}&=-2K\Phi^2\sin\psi+\xi''.
\end{align}

When $\Phi$ is treated as a parameter, the probability distribution of $\psi$, $P(\psi,t)$, is effectively governed by the following Fokker-Planck equation:
\begin{align}
 \frac{\partial P}{\partial t}=2K\Phi^2\frac{\partial}{\partial \psi}(\sin\psi P) +
D\frac{\partial^2 P}{\partial \psi^2}. \label{eq:fokker}
\end{align}
A curve in $\Phi$-$\psi$ space is defined by equating the drift term and the diffusion term, while replacing the
derivative of $\psi$ by $\psi$ itself:
 \begin{align}
K\Phi^2\psi\sin\psi = D. \label{eq:drift_diffusion}
 \end{align}
 Note that here and in the subsequent analysis, we only focus on the scaling form, disregarding numerical factors. 
Hence, if $\Phi$ satisfies 
\begin{align}
 \Phi \lesssim \Phi_c \equiv \sqrt{\frac{D}{K}}, \label{eq:diffusion_cond}
\end{align}
then the effect of the drift term is dominated by the diffusion term for the whole
range of $\psi$. Conversely, if $\Phi>\Phi_c$, the drift term becomes effective. If $\Phi$ evolves as $\Phi=\omega t$, we can determine the critical time $t_c$ at
which $\Phi$ reaches $\Phi_c$:
\begin{align}
 t_c = \frac{1}{\omega}\sqrt{\frac{D}{K}}.
\end{align}

Now consider a trajectory starting from $(\Phi,\psi)=(0,0)$ at $t=0$. The typical diffusion length
of $\psi$ at $t=t_c$ is given by
\begin{align}
 \psi_1 = \sqrt{Dt_c} = \omega^{-\frac{1}{2}}D^{\frac{3}{4}}K^{-\frac{1}{4}}.
\end{align}

Moreover, $\dot{\Phi}=0$ with $D=0$ determines another curve in $\Phi$-$\psi$ space:
\begin{align}
K\Phi\sin^2\frac{\psi}{2} = \omega. \label{eq:phi0}
\end{align}
For large $K$, this curve and the other curve given by Eq.~\eqref{eq:drift_diffusion} cross each other, where $\psi=\psi_2$ at the intersection is shown to
be small. Indeed, by expanding Eqs.~\eqref{eq:drift_diffusion} and \eqref{eq:phi0} in terms of $\psi$,
$\psi_2$ is obtained as
\begin{align}
 \psi_2 = \omega^{\frac{1}{2}}D^{-\frac{1}{2}}K^{-\frac{1}{2}}, 
\end{align}
which diminishes as $K$ increases. In order that the trajectory reaches the curve \eqref{eq:phi0} before
it gets affected by the drift term, the diffusion length $\psi_1$ must be greater than $\psi_2$. The
critical transition line is given by $\psi_1=\psi_2$, which yields the scaling form
\begin{align}
 D = \omega^{\frac{4}{5}}K^{-\frac{1}{5}}, \label{eq:scaling}
\end{align}
which implies that, for a fixed $D$ value, phase slip becomes more and more frequent as $K$
increases.
This scaling relation is in good agreement with the numerically obtained reentrant transition line in
Fig.~\ref{fig:phasediag_mutual}(b). 

As mentioned above, the unilateral coupling case shows a different scaling relation
$D=\omega^{\frac{4}{3}}K^{-\frac{1}{3}}$, which indicates that the reentrant region is narrower in
the present case than in the unilateral case. 

\subsection{Oscillation death}\label{sec:death}
In the noise-free case, in addition to the synchronous solution $(\Phi,\psi)=(\omega t, 0)$, the system
may have steady state solutions, which correspond to oscillation death, depending on the parameters $K$ and $\alpha$. Here, without loss of
generality, we restrict the
range of $\Phi$ and $\psi$ to $0\le\Phi<\pi$ and $0\le\psi<2\pi$, respectively. Steady
states are given by $\dot{\Phi}=0$ and $\dot{\psi}=0$ in Eqs.~\eqref{eq:phi_model} and
\eqref{eq:psi_model} with $D=0$, which are satisfied by $(\Phi, \psi)=(\Phi^*,
\pi)$ and $(\Phi, \psi)=(\alpha, \psi^*)$, where $\Phi^*$ and $\psi^*$ are determined by
\begin{align}
 \omega - K\{\sin(2\Phi^*-\alpha)+\sin\alpha\}=0, \label{eq:phi}
\end{align}
and 
\begin{align}
 \omega - 2K\sin\alpha\sin^2\frac{\psi^*}{2}=0, \label{eq:psi}
\end{align}
respectively. 

The solution $\Phi^*$ to Eq.~\eqref{eq:phi} exists only when inequality
\eqref{eq:kc1} is satisfied. At $K=K_{c1}$, two solutions $\Phi^*=\Phi_a$ and $\Phi^*=\Phi_b$
appear as a result of saddle-node bifurcation, where
\begin{align}
 \Phi_a = \frac{\alpha}{2}+\frac{\theta}{2}, \quad \Phi_b=\frac{\alpha}{2}+\frac{\pi-\theta}{2}, 
\end{align}
and $\theta=\arcsin(\frac{\omega}{K}-\sin\alpha)$. Linear stability analysis shows that at the
onset of bifurcation $\Phi^*=\Phi_a$ and $\Phi^*=\Phi_b$ correspond to a saddle and an unstable
focus, respectively. The unstable eigenvalue for the saddle is given by
$\lambda=K(\cos\alpha-\sqrt{1-(\frac{\omega}{K}-\sin\alpha)^2})$. Hence, at $K=K_{c2} > K_{c1}$, secondary bifurcation occurs, where $K_{c2}$ is determined by $\lambda=0$.
It is easy to see that if $\alpha\le 0$, then there is no such $K_{c2}$ that satisfies $\lambda=0$. 
If $\alpha>0$, this yields
\begin{align}
 K_{c2} = \frac{\omega}{2\sin\alpha}.\label{eq:kc2}
\end{align}

For $K>K_{c2}$ and $\alpha>0$, Eq.~\eqref{eq:psi} also has two solutions $\psi^*=\beta$ and
$\psi^*=2\pi-\beta$, where $\beta=2\arcsin\sqrt{\frac{\omega}{2K\sin\alpha}}$.
Linear stability
analysis shows that these two branches are saddles, which originate from $(\Phi,\psi)=(\Phi_a,\pi)$
at $K=K_{c2}$ by pitchfork bifurcation. Note that $\theta=\alpha$ and $\beta=\pi$ at $K=K_{c2}$, and therefore $(\Phi_a,
\pi)=(\alpha,\psi^*)$ at this point. At $K>K_{c2}$, point $(\Phi_a, \pi)$ becomes a stable
focus, which corresponds to the death state. 

\begin{figure}[htb]
\includegraphics[width=.7\linewidth]{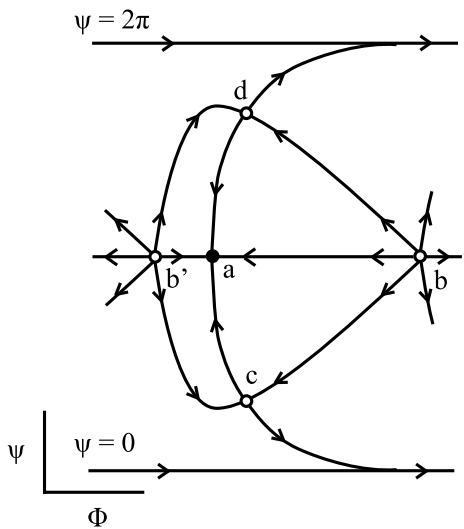}
 \caption{Schematic of the $\Phi$-$\psi$ phase portrait for Eqs.~\eqref{eq:phi} and \eqref{eq:psi}
 for $\alpha>0$ and $K > K_{c2}$. Fixed
 points, periodic orbits, and invariant manifolds are drawn. The points refer to: (a) $(\Phi,\psi)=(\Phi_a, \pi)$, (b) $(\Phi,\psi)=(\Phi_b,\pi)$, (b') $(\Phi,\psi)=(\Phi_b-\pi,\pi)$ (c) $(\Phi,\psi)=(\alpha,\beta)$, and
 (d) $(\Phi,\psi)=(\alpha,\pi-\beta)$. Note that $b'$ is equivalent to $b$. }\label{fig:death_schematic}
\end{figure}

Thus we find that our system can undergo oscillation death if $\alpha>0$ and $K>K_{c2}$.
If $\Phi$ is stretched into the range $(-\infty, \infty)$, the death states are located on
$(\Phi,\psi)=(\Phi_a+2\pi n, \pi)$ ($n \in \mathbb{Z}$) and are separated from the synchronous
solution by separatrices, which are invariant manifolds connecting the unstable foci
$(\Phi,\psi)=(\Phi_b+2\pi(n-1),\pi)$ and $(\Phi_b+2\pi n,\pi)$ and the saddles
$(\Phi,\psi)=(\alpha+2\pi n, \beta)$ and $(\alpha+2\pi n, 2\pi-\beta)$ (see Fig.~\ref{fig:death_schematic}). These separatrices can be
overcome when a sufficiently strong noise is applied to the system.

It is possible that the conditions for nontrivial phase slip and oscillation death are both satisfied. Since nontrivial phase slip occurs along $\dot{\Phi}=0$, the trajectory is likely to be
trapped during a slip event by the death state $(\Phi_a, \pi)$, which is located on the line
$\dot{\Phi}=0$. Then, the probability of trapping depends on the noise intensity, as indicated in
Fig.~\ref{fig:death}: for a fixed $K > K_{c2}$, strong noise aids escape from the death state; conversely, weak noise is not sufficient to escape from the synchronous state. Thus the boundary
of the death state moves rightward for large and small $D$ values. 

\section{Network-connected system}\label{sec:network}
Now we consider $N$ coupled phase oscillators with frequency $\omega$ under noise. The $i$-th oscillator obeys
\begin{equation}
 \dot{\phi}_i = \omega + K Z(\phi_i)\sum_{j=1}^N A_{ij}\left\{h(\phi_j)-h(\phi_i)\right\}+\xi_i,\label{eq:network_model}
\end{equation}
where $Z$ and $h$ are given by Eq.~\eqref{eq:functions}, and $\xi_i$ is the same as before. Their
connections are determined by adjacency matrix $A$. 
We investigate synchronizability of the following networks: scale-free networks generated by the
Barab\'{a}si-Albert algorithm \cite{albert02rmp} with the minimum degree $m_0=1$ with network size
$N=100$ (average degree $\av{d}=2.0$) or
$N=10000$ ($\av{d}=2.0$), and $m_0=3$ with
$N=10000$ ($\av{d}=6.0$); all-to-all connection with $N=100$ ($\av{d}=99$); and an
Erd\H{o}s-R\'{e}nyi random network with $N=400$ ($\av{d}=5.8$).
Since we are interested in destabilization of the synchronous state, we start with a
synchronous initial condition with weak random perturbations in the range $(-0.01\pi,0.01\pi)$ given
to individual oscillators.

For a given network, a phase slip event of the $i$th node is counted when a full revolution of $\phi_i$ in
the positive or negative direction is made with respect to the mean phase of the rest of the oscillators: the phase difference for $i$ is defined as $\psi_j \equiv \phi_j - \frac{1}{N-1}\sum_{i\neq j}\phi_i =
\frac{N}{N-1}(\phi_j - \Phi_N)$, where $\Phi_N \equiv \frac{1}{N}\sum_{j=1}^N\phi_j$ is the mean
phase of all oscillators. The phase slip rate of oscillator $i$ is then the total number of
phase slip events divided by the observation time $t_{\mathrm{max}}$, which in the following simulations is $t_{\mathrm{max}}=10^4$. 

\begin{figure}[htb]
\includegraphics[width=\linewidth]{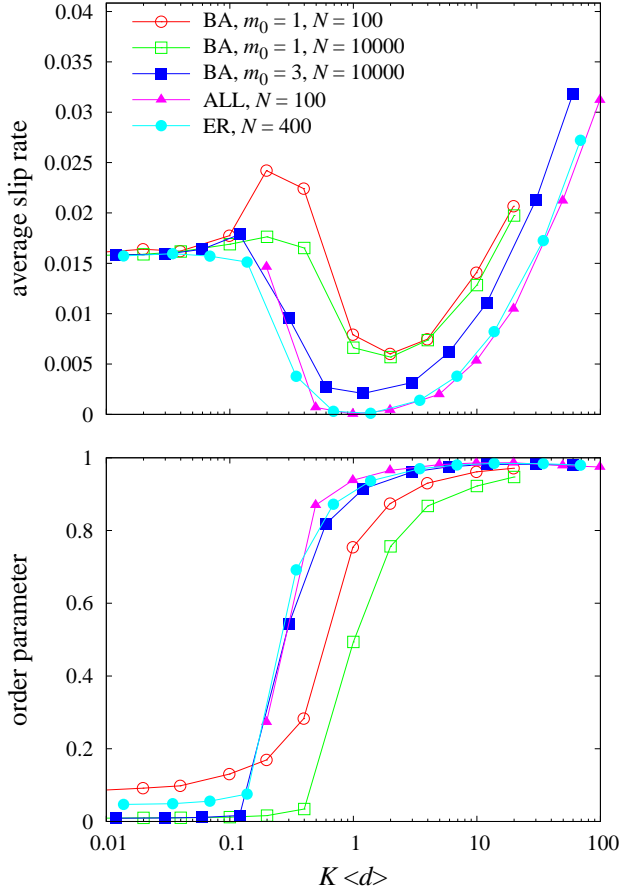}
 \caption{Average phase slip rate (top) and order parameter (bottom) for
 Eq.~\eqref{eq:network_model} with $\alpha=0$ and $D=0.1$ as a function of the effective coupling strength $
 K\langle d \rangle$. Different kinds of networks are employed: Barab\'{a}si-Albert scale-free
 networks (BA) with different values of the minimum degree $m_0$ and network size $N$; an
 all-to-all network (ALL); and an Erd\H{o}s-R\'{e}nyi random network (ER). }\label{fig:slip_a0.0}
\end{figure}

\subsection{Average phase slip rate}
The average phase slip rates over all oscillators are shown in Fig.~\ref{fig:slip_a0.0} for
different networks with a nonzero value of the noise intensity $D=0.1$,
where the horizontal axis is the effective coupling strength $K\av{d}$. All networks show the same
dependency, namely that as the coupling strength increases, the slip rate first starts to drop and then
increases. However, to what extent the slip rate drops differs: the scale-free networks show higher
slip rates than all-to-all or random networks. Also, the two scale-free networks with the
minimum degree $m_0=1$ shows a higher slip rate than that with $m_0=3$. For the scale-free networks,
network size, i.e., $N=100$ or $N=10000$, does not make a noticeable difference.
In contrast to the slip rate, the order parameter $\sigma$ defined by
$\sigma=|\frac{1}{N}\sum_{i=1}^N e^{i\phi_i}|$ shows similar dependence on $K\av{d}$ for different
networks: the order parameter increases with $K$ and remains high even when the slip rate
is high. 

A slight increase of the slip rate in the low-$K$ region is caused by incoherent input to the
high-degree nodes: if a high-degree node $i$ with degree $d_i$ receives many incoherent inputs, the
sum of all inputs in Eq.~\eqref{eq:network_model} is $\sum_i
h(\phi_i)\sim 0$, and thus its dynamics are effectively governed by
\begin{align}
 \dot{\phi}_i   &= \omega + Kd_i \sin(\phi_i-\alpha)\cos\phi_i + \xi_i. 
\end{align}
When noise is absent, the average period $T$ is given by
\begin{align}
 T&=\int_0^{2\pi}\frac{d\phi}{\omega + Kd_i\sin(\phi-\alpha)\cos\phi}\nonumber\\
 &=  \frac{2\pi}{\omega}\left[1-\frac{Kd_i}{\omega}\sin\alpha-\frac{K^2d_i^2}{4\omega^2}\cos^2\alpha\right]^{-\frac{1}{2}}.\label{eq:period_divergence}
\end{align}
For $\alpha=0$, the oscillation period increases as $K$ increases from zero, and diverges at $K=2\omega/d_i$. If the connected nodes start to synchronize, it starts to
oscillate again. As indicated in Fig. \ref{fig:slip_a0.0}, this tends to occur in scale-free
networks: when the coupling strength is weak, the low-degree nodes are incoherent, and the
high-degree nodes receive a lot of such incoherent signals. 

\begin{figure}[hbtp]
\includegraphics[width=\linewidth]{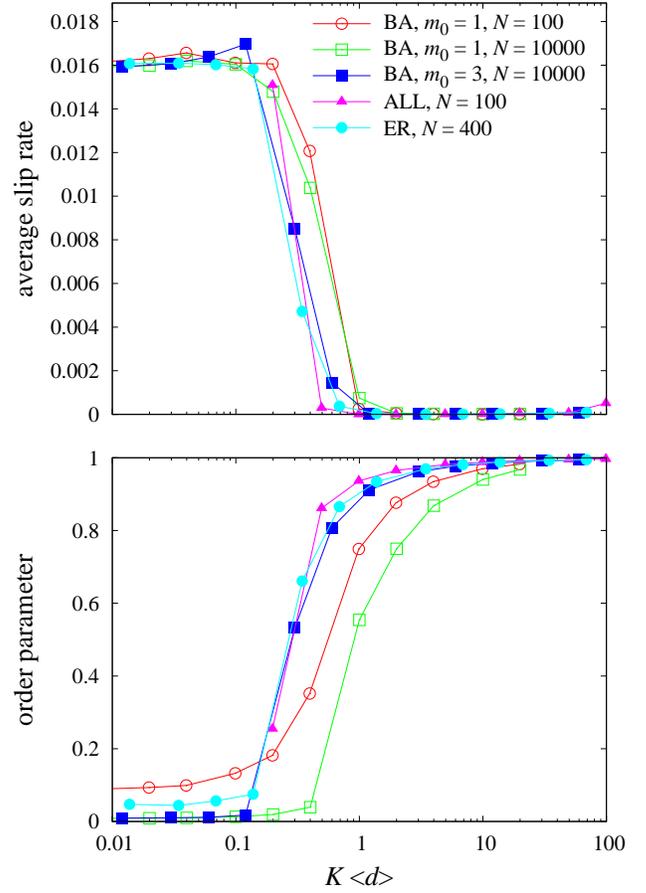}
 \caption{Same as Fig.~\ref{fig:slip_a0.0}, except with $\alpha=-0.3$. }\label{fig:slip_a-0.3}
\end{figure}

\subsection{Dependence on the model parameter}
The situation drastically changes when the model parameter is chosen to be $\alpha=-0.3$, as shown
in Fig.~\ref{fig:slip_a-0.3}: in this case, there are no noticeable differences among the different
networks. No nontrivial phase slip is observed in any network, while the order parameter behaves in
the same way as in the case of $\alpha=0$, which is consistent with the analysis for the mutual coupling
case. This indicates that different behavior among the networks results from the existence of the reentrant region. 

\begin{figure}[tbhp]
\includegraphics[width=\linewidth]{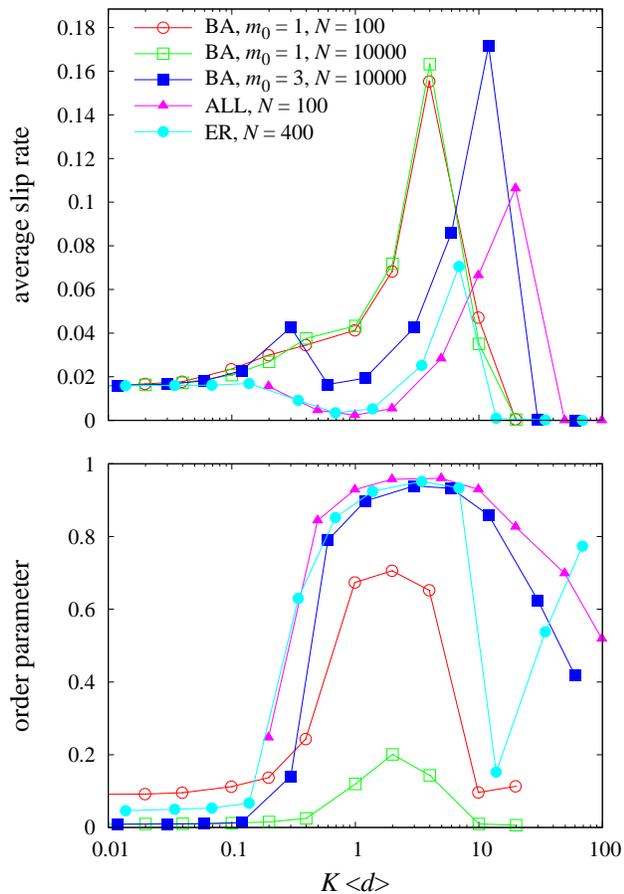}
 \caption{Same as Fig.~\ref{fig:slip_a0.0}, except with $\alpha=0.3$. }\label{fig:slip_a0.3}
\end{figure}

Conversely, the nontrivial phase slip is more likely to occur for $\alpha=0.3$ (Fig.~\ref{fig:slip_a0.3}). The
average slip rate increases especially when the network is scale-free, where a drop in the middle
range of $K$ is no longer observed, as in the case of $\alpha=0$, and owing to frequent phase slip events,
the order parameter is also suppressed. The random and the all-to-all networks show slight
differences when the average slip rate starts to increase at $K\av{d}>1$, which is not obvious for
$\alpha=0$. The all-to-all network shows lower slip rates.
As $K$ increases further, a sudden drop of the average slip rate is observed for all networks,
indicating oscillation death. As a result of oscillation death, oscillators fall into a two-cluster
state with the phase difference between the clusters depending on the network structure, which is
reflected in the $\sigma$ values for individual oscillators in the death state.
Again, the all-to-all network is more resistant
to oscillation death, compared with the random network, indicating that the all-to-all network has
higher synchronizability. 

\subsection{Dependence on the noise intensity}
We have checked in our preliminary numerical analysis that when $D=0$ the system converges to the
in-phase state and no phase slip is observed for all five networks with the
three $\alpha$ values: $\alpha=-0.3$, $0$, and $0.3$. This is in accordance with the observation in
the case of two oscillators, where a nonzero amount of noise is crucial for phase slip to occur. 

Note that, if noise is absent and the averaging is valid, Eq.~\eqref{eq:network_model} with
scale-free networks reduces to the model studied in Ko and Ermentrout\cite{ko_ermentrout}, where partial
synchronization has been observed when asymmetry of the coupling ($\alpha$ in our model) is
sufficiently large. In contrast to this, we have not observed partial synchronization for $D=0$, even when we start with fully random phases. This is presumably because our $\alpha$ values are not sufficiently large. 

\begin{figure}[tbh]
\includegraphics[width=\linewidth]{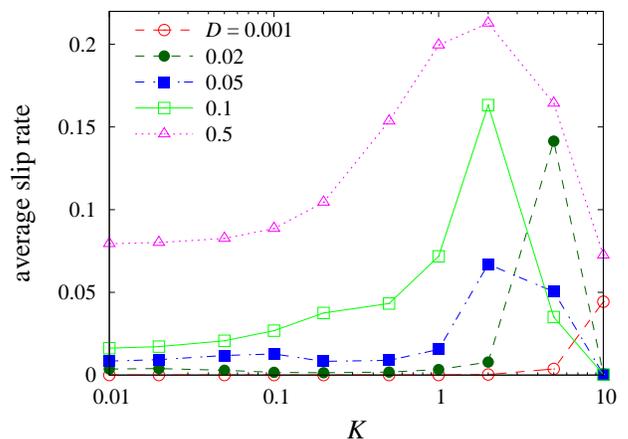}
 \caption{Noise dependence of the average slip rate of a
 scale-free network with $m_0=1$, $N=10000$ and $\alpha=0.3$, as a function of $K$.}\label{fig:noise_dependence}
\end{figure}

We have also calculated the phase slip rate for a nonzero but small value of the noise intensity $D=0.01$. In this
case, for all five networks studied above and for all three values of $\alpha$, we have not observed
nontrivial phase slip except for the BA model with $m_0=1$ and $\alpha=0.3$, for which we have further
investigated the noise intensity dependence (Fig~\ref{fig:noise_dependence}). For $K\le 2$, there is
a clear dependency, namely that the average slip rate increases as the noise intensity increases. This
indicates that nontrivial phase slip is enhanced by noise, which is in accordance with
the two-oscillator case (Fig.~\ref{fig:phasediag_mutual}). On the other
hand, for $K>2$ the average slip rate does not show monotonous dependence on $D$ since partial
oscillation death occurs for smaller $D$ values, which increases slips between active and inactive
nodes.

\subsection{Average frequency of individual oscillators}
\begin{figure}[tbh]
\includegraphics[width=\linewidth]{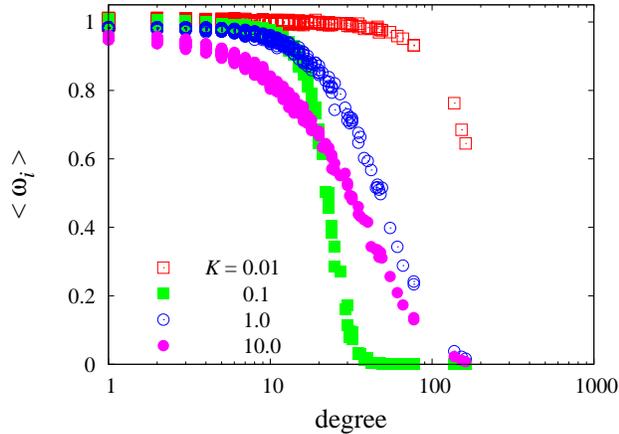}
 \caption{Average frequency distribution of a scale-free network with $m_0=1$, $N=10000$, and $\alpha=0$. }\label{fig:distrib_bamin1}
\end{figure}

The slip rate of individual oscillators depends on their degree. Figure~\ref{fig:distrib_bamin1} plots
the time-averaged frequency of individual oscillators, $\av{\omega_i}$, for the scale-free network
with $m_0=1$ and $N=10000$, which shows that high-degree nodes have smaller frequencies. Even when $K$ is small,
the frequencies of the highest degree nodes are already small, and at $K=0.1$ nodes with degree $d> 40$ halt, indicating the period divergence
described in Eq.~\eqref{eq:period_divergence}. At $K=1.0$, high-degree nodes start to increase their
frequencies again, by synchronizing with low-degree nodes, whereas low-degree nodes in turn
decreases in frequency. Note that $K=1.0$ corresponds to the lowest average slip rate
(Fig.~\ref{fig:slip_a0.0}, $K\av{d}=2$). The frequency drops entirely as $K$ increases further. In
this way, high-degree nodes always show low frequencies, either because of incoherent inputs or nontrivial phase slip. 

\begin{figure}[tbh]
\includegraphics[width=\linewidth]{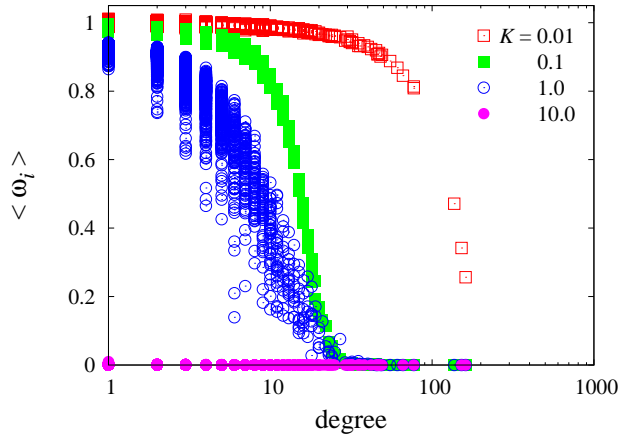}
 \caption{Average frequency distribution of a scale-free network with $m_0=1$, $N=10000$, and $\alpha=0.3$. }\label{fig:distrib_bamin1a0.3}
\end{figure}

For $\alpha=0.3$, the frequency decreases over the entire range of degrees as $K$ increases, as shown
in Fig.~\ref{fig:distrib_bamin1a0.3}. Also, at $K=1.0$, a wide distribution of frequencies is
observed in nodes with the same degrees. A complete oscillation death is observed at $K=10.0$. 

Nontrivial phase slip of the high-degree nodes can be understood as follows: suppose that the entire network is
close to synchronization with order parameter $\sigma \sim 1$; since a high-degree node receives
coherent inputs from many other nodes, Eq.~\eqref{eq:network_model} for a high-degree node $i$ with
degree $d_i$ is
approximated by
\begin{align}
 \dot{\phi}_i = \omega + Kd_iZ(\phi_i)\{h(\av{\Phi})-h(\phi_i)\}+\xi_i,
\end{align}
where $\av{\Phi}\sim \omega t$ is the mean phase of the nodes connected to $i$. Thus the
situation is almost the same as for the oscillator-pacemaker system \eqref{eq:previous}, where the coupling
strength $K$ is effectively enhanced by its degree $d_i$. 
Since scale-free networks have a large
heterogeneity in degree distribution, differences of the effective coupling constant become
significant. Therefore, if $\alpha\ge 0$, that is, if the model has a reentrant
transition, then there can be a situation where low-degree nodes are located in the region of
incoherence whereas high-degree nodes are in the reentrant region. In this case synchronizability
as a whole will be considerably reduced.

\section{Conclusions}
We have studied networks of phase oscillators under the influence of noise, where the coupling can be so strong that
averaging is not necessarily valid, and found that networks show different synchronization
properties when measured by their phase slip rates.  We have also observed that heterogeneous
networks suffer more strongly from nontrivial
phase slip in the strong coupling regime. This nontrivial phase slip is understood from the phase
diagram of two coupled phase oscillators, where the coupling can be unilateral or mutual. In both
cases, there is a range of model parameters in which a reentrant transition to the nontrivial phase
slip state is observed. It is this reentrant property that underlies poor synchronizability of
heterogeneous networks, where nodes with both low and high degrees are likely to be out of the synchronization
range. 

\section*{Acknowledgment}
H. K. acknowledges financial support from CREST, JST.

\end{document}